\documentclass[onecolumn]{revtex4}
\usepackage{graphicx}
\usepackage[dvips]{color}
\usepackage[english]{babel}
\newcommand{\gcmt}{{\rm g/cm}^3}
\newcommand{\msun}{M_{\odot}}
\newcommand{\fm}{{\rm fm}}
\newcommand{\ergs}{{\rm erg/s}}
\newcommand{\lsim}{\stackrel{\textstyle <}{_\sim}}
\newcommand{\gsim}{\stackrel{\textstyle >}{_\sim}}

\begin{document}

\markboth{F.~Weber, G.~A.~Contrera, M.~Orsaria, W.~Spinella, O.~Zubairi}
{Properties of High-Density Matter in Neutron Stars}

\title{PROPERTIES OF HIGH-DENSITY MATTER IN NEUTRON STARS}

\author{Fridolin Weber} \email{fweber@mail.sdsu.edu}
\affiliation{Department of Physics, San Diego State University, 5500
             Campanile Drive, San Diego, California 92182, USA}
\affiliation{Center for Astrophysics and Space Sciences, University of California,\\
             San Diego, La Jolla, California 92093, USA.}

\author{Gustavo A. Contrera} \email{contrera@fisica.unlp.edu.ar}
\affiliation{Department of Physics, San Diego State University, 5500
             Campanile Drive, San Diego, California 92182, USA,}
\affiliation{CONICET, Rivadavia 1917, 1033 Buenos Aires, Argentina,}
\affiliation{IFLP, CONICET - Dpto. de F{\'i}sica, UNLP, La Plata, Argentina}
\affiliation{Gravitation, Astrophysics and Cosmology Group, Facultad de Ciencias Astron{\'o}micas y
             Geof{\'i}sicas, UNLP, Paseo del Bosque S/N (1900), La Plata, Argentina.}

\author{Milva G. Orsaria} \email{morsaria@rohan.sdsu.edu}
\affiliation{Department of Physics, San Diego State University, 5500
             Campanile Drive, San Diego, California 92182, USA,}
\affiliation{CONICET, Rivadavia 1917, 1033 Buenos Aires, Argentina}
\affiliation{Gravitation, Astrophysics and Cosmology Group, Facultad de Ciencias Astron{\'o}micas y
             Geof{\'i}sicas, UNLP, Paseo del Bosque S/N (1900), La Plata, Argentina.}

\author{William Spinella} \email{spinellla@mail.sdsu.edu}
\affiliation{Department of Physics \& Computational Science Researech Center, San Diego State
             University, 5500 Campanile Drive, San Diego, California 92182, USA.}

\author{Omair Zubairi} \email{zubairi@rohan.sdsu.edu}
\affiliation{Department of Physics \& Computational Science Researech Center, San Diego State
             University, 5500 Campanile Drive, San Diego, California 92182, USA.}

\begin{abstract}
This short review aims at giving a brief overview of the various
states of matter that have been suggested to exist in the ultra-dense
centers of neutron stars. Particular emphasis is put on the role of
quark deconfinement in neutron stars and on the possible existence of
compact stars made of absolutely stable strange quark matter (strange
stars).  Astrophysical phenomena, which distinguish neutron stars from
quark stars, are discussed and the question of whether or not quark
deconfinement may occur in  neutron stars is
investigated.
Combined with observed astrophysical data, such studies
are invaluable to delineate the complex structure of
compressed baryonic matter and to put firm constraints on the largely
unknown equation of state of such matter.

\end{abstract}

\date{\today}


\maketitle

\section{Introduction}

Exploring the properties of compressed baryonic matter, or, more
generally, strongly interacting matter at high densities and/or
temperatures, has become a forefront area of modern physics.
Experimentally, the properties of such matter are being probed with
the Relativistic Heavy Ion Collider RHIC at Brookhaven and the Large
Hadron Collider (LHC at CERN). Great advances in our understanding of
such matter are expected from the next generation of heavy-ion
collision experiments at FAIR (Facility for Antiproton and Ion
Research at GSI) and NICA (Nuclotron-bases Ion Collider fAcility at
JINR) \cite{CBMbook,NICA,Klaehn12:a}.

Neutron stars contain compressed baryonic matter permanently in their
centers.  Such stars are remnants of massive stars that blew apart in
core-collapse supernova explosions. They are typically about 20
kilometers across and spin rapidly, often making many hundred
rotations per second.  Many neutron stars form radio pulsars, emitting
radio waves that appear from the Earth to pulse on and off like a
\begin{figure}[htb]
\centering
\includegraphics[width=0.7 \textwidth]{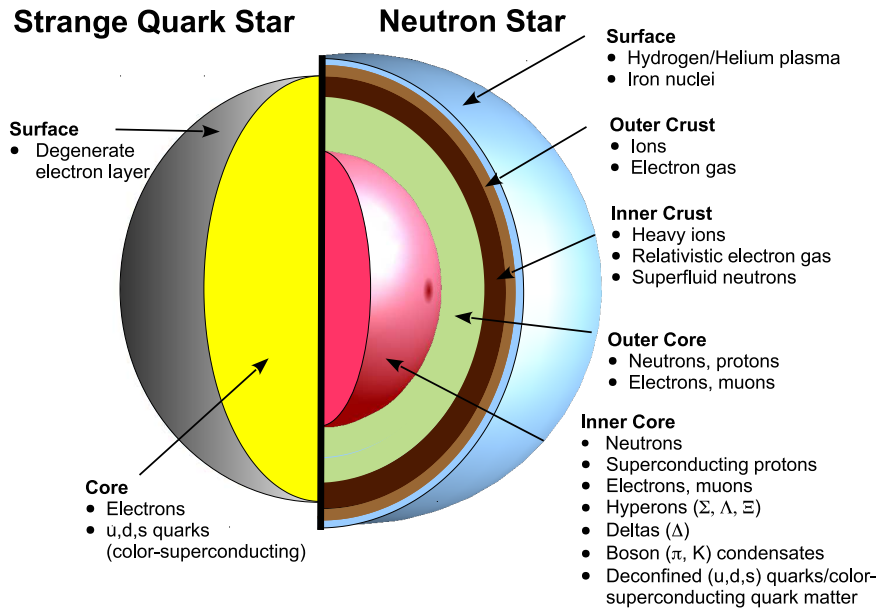}
\caption{(Color online) Schematic structures of quark stars (left) and
  neutron stars (right). If the strange quark matter hypothesis should
  be correct, most, if not all, neutron stars should in fact be strange
  quark stars.}
\label{fig:qs-ns}
\end{figure}
lighthouse beacon as the star rotates at very high speeds. Neutron
stars in X-ray binaries accrete material from a companion star and
flare to life with tremendous bursts of X-rays. Depending on mass and
rotational frequency, gravity compresses the matter in the cores of
neutron stars to densities that are several times higher than the
density of ordinary atomic nuclei.  At such huge densities atoms
themselves collapse, and atomic nuclei are squeezed so tightly
together that new fundamental particles may be produced and novel
states of matter be created.  The most spectacular phenomena stretch
from the generation of hyperons and delta particles, to the formation
of meson condensates, to the formation of a plasma of deconfined
quarks (see Fig.\ \ref{fig:qs-ns}).  The interest in the role of quark
deconfinement for astrophysics has received renewed interest by the
discovery that quark matter at low temperatures ought to be a color
superconductor (see Ref.\ \cite{alford08:a} and references
therein).

All these features make neutron stars superb astrophysical
laboratories for a wide range of physical studies \cite{glen97:book,%
frido99:book,frido:review05,lattimer01:a,blaschke01:trento,sedrakian07:a,%
page06:review,hell14:a}.  And with
observational data accumulating rapidly from both orbiting and ground
based observatories spanning the spectrum from X-rays to radio
wavelengths, there has never been a more exiting time than today to
study neutron stars and associated catastrophic astrophysical events.
The Hubble Space Telescope and X-ray satellites such as Chandra and
XMM-Newton in particular have proven especially valuable.  New
astrophysical instruments such as the Five hundred meter Aperture
Spherical
\begin{table}[htb]
\caption{Theoretical properties of strange quark stars and neutron stars
      compared.}
{\begin{tabular}{@{}ll@{}}\toprule
Strange Quark Stars   &Neutron Stars  \\ \colrule
Made entirely of deconfined up, down, strange      &Nucleons, hyperons, boson condensates, \\
~~quarks, and electrons                              &deconfined quarks, electrons, and muons \\
Absent                                             &Superfluid neutrons \\
Absent                                             &Superconducting protons \\
Color superconducting quarks                       &Color superconducting quarks \\
Energy per baryon $\lsim 930$~MeV                  &Energy per baryon $> 930$~MeV \\
Self-bound ($M \propto R^3$)                       &Bound by gravity \\
Maximum mass $\sim 2 \, \msun$                     &Same \\
No minimum mass if bare                            &Minimum mass $\sim 0.1 \, \msun$ \\
Radii $R\lsim 10- 12 $~km                          &Radii $R \gsim 10 - 12$~km \\
Baryon numbers $B \lsim 10^{57}$                   &Baryon numbers $10^{56}  \lsim B \lsim 10^{57}$ \\
Electric surface fields $\sim 10^{18}$ to $\sim 10^{19}$~V/cm     &Absent   \\
Can either be bare or enveloped in thin            &Always have nuclear crusts \\
~~nuclear crusts (masses $\lsim 10^{-5}\, \msun$)  &    \\
Maximum density of crust set by neutron drip, i.e.,  &Does not apply, i.e., neutron stars   \\
~~strange stars posses only outer crusts            &~~posses inner and outer crusts \\
Form two-parameter stellar sequences              &Form one-parameter stellar sequences \\ \botrule
\end{tabular}  \label{tab:comparison}}
\end{table}
Telescope (FAST), the square kilometer Array (skA), Fermi Gamma-ray
Space Telescope (formerly GLAST), Astrosat, ATHENA (Advanced Telescope
for High ENergy Astrophysics), and the Neutron Star Interior
Composition Explorer (NICER) promise the discovery of tens of
thousands of new neutron stars. Of particular interest will be the
proposed NICER mission (scheduled to launch in 2016), which is
dedicated to the study of extraordinary gravitational,
electromagnetic, and nuclear-physics environments embodied by neutron
stars. NICER will explore the exotic states of matter in the core
regions of neutron stars, where, as mentioned just above, density and
pressure are considerably higher than in atomic nuclei, confronting
nuclear theory with unique observational constraints.

There is also the theoretical possibility that quark matter made of
up, down and strange quarks (so-called strange quark
matter \cite{farhi84:a}) may be more stable than ordinary nuclear
matter \cite{witten84:a}.  This so-called strange matter hypothesis
constitutes one of the most startling possibilities regarding the
behavior of superdense matter, which, if true, would have implications
of fundamental importance for cosmology, the early universe, its
evolution to the present day, and astrophysical compact objects such
as neutron stars and white dwarfs (see Ref.\ \cite{weber05:a} and
references therein). The properties of compact stars made of strange
quark matter, referred to as strange (quark) stars, are compared with
those of neutron stars in Table~\ref{tab:comparison} and
Fig.\ \ref{fig:qs-ns}.  Even after three decades of research, there is
no sound scientific basis on which one can either confirm or reject
the strange quark matter hypothesis so that it remains a serious
possibility of fundamental significance for various astrophysical
phenomena, as discussed in the next section \cite{EMMI14}.

\section{Properties of strange quark stars}

A bare quark star differs qualitatively from a neutron star which has
a density at the surface of about 0.1 to $1~\gcmt$. The thickness of
the quark surface is just $\sim 1~\fm$, the length scale of the strong
interaction. The electrons at the surface of a quark star are held to
quark matter electrostatically, and the thickness of the electron
surface is several hundred fermis.  Since neither component, electrons
and quark matter, is held in place gravitationally, the Eddington
limit to the luminosity that a static surface may emit does not apply,
so that bare quark stars may have photon luminosities much greater
than $10^{38}~\ergs$.  It has been shown in Ref.\ \cite{usov98:a} that this
value may be exceeded by many orders of magnitude by the luminosity of
$e^+ e^-$ pairs produced by the Coulomb barrier at the surface of a
hot strange star. For a surface temperature of $\sim 10^{11}$~K, the
luminosity in the outflowing pair plasma was calculated to be as high
as $\sim 3 \times 10^{51}~\ergs$.  Such an effect may be a good
observational signature of bare strange
stars \cite{usov98:a,usov01:c,usov01:b,cheng03:a}. If the strange star
is dressed, that is, enveloped in a nuclear crust, however, the
surface made of ordinary atomic matter would be subject to the
Eddington limit. Hence the photon emissivity of a dressed quark star
would be the same as for an ordinary neutron star.  If quark matter at
the stellar surface is in the CFL (Color-Flavor Locked) phase the
process of $e^+ e^-$ pair creation at the stellar quark matter surface
may be turned off.  This may be different for the early stages of a
very hot CFL quark star \cite{vogt03:a}.

In contrast to neutron stars, the radii of self-bound quark stars
decrease the lighter the stars, according to $M \propto R^3$. The
existence of nuclear crusts on quark stars changes the situation
drastically \cite{weber05:a,weber93:b}.  Since the crust is bound
gravitationally, the mass-radius relationship of quark stars with
crusts can be qualitatively similar to mass-radius relationships of
neutron stars and white dwarfs \cite{weber93:b}.  In general, quark
stars with or without nuclear crusts possess smaller radii than
neutron stars. This implies that quark stars have smaller mass
shedding (break-up) periods than neutron stars. Moreover, due to the
smaller radii of quarks stars, the complete sequence of quark
stars--and not just those close to the mass peak, as it is the case
for neutron stars--can sustain extremely rapid
rotation \cite{weber05:a,weber93:b}.  In particular, a strange star
with a typical pulsar mass of around $1.45\,\msun$ has a Kepler period
in the approximate range of $0.55\lesssim{P_{\rm K}}/{\rm
  msec}\lesssim 0.8$ \cite{weber93:b,glen92:crust}. This is to be
compared with ${P_{\rm K}} \sim 1~{\rm msec}$ for neutron stars of the
same mass.

One of the most amazing features of strange quark stars concerns the
existence of ultra-high electric fields on their surfaces, which, for
ordinary (i.e., non-superconducting) quark matter, is around
$10^{18}$~V/cm.  If strange matter forms a color superconductor, as
expected for such matter, the strength of the electric field may
increase to values that exceed $10^{19}$~V/cm. The energy density
associated with such huge electric fields is on the same order of
magnitude as the energy density of strange matter itself, which may alter
the masses and radii of strange quark stars at the 15\% and 5\% level,
respectively \cite{negreiros09:a}.

The electrons at the surface of a quark star are not necessarily in a
fixed position but may rotate with respect to the quark matter
star \cite{negreiros10:a}. In this event magnetic fields can be
generated which, for moderate effective rotational frequencies between
the electron layer and the stellar body, agree with the magnetic
fields inferred for several Central Compact Objects (CCOs). These
objects could thus be interpreted as quark stars whose electron
atmospheres rotate at frequencies that are moderately different ($\sim
10$~Hz) from the rotational frequency of the quark star itself.

Last but not least, we mention that the electron surface layer may be
strongly affected by the magnetic field of a quark star in such a way
that the electron layer performs vortex hydrodynamical
oscillations \cite{xu12:a}. The frequency spectrum of these
oscillations has been derived in analytic form in
Ref.\ \cite{xu12:a}. If the thermal X-ray spectra of quark stars
are modulated by vortex hydrodynamical oscillations, the thermal
spectra of compact stars, foremost central compact objects (CCOs) and
X-ray dim isolated neutron stars (XDINSs), could be used to verify the
existence of these vibrational modes observationally. The central
compact object 1E 1207.4-5209 appears particularly interesting in this
context, since its absorption features at 0.7 keV and 1.4 keV can be
comfortably explained in the framework of the hydro-cyclotron
oscillation model \cite{xu12:a}.

Rotating superconducting quark stars ought to be threaded with
rotational vortex lines, within which the star's interior magnetic
field is at least partially confined.  The vortices (and thus magnetic
flux) would be expelled from the star during stellar spin-down,
leading to magnetic re-connection at the surface of the star and the
prolific production of thermal energy. It has been shown in
Ref.\ \cite{niebergal10:a} that this energy release can re-heat
quark stars to exceptionally high temperatures, such as observed for
Soft Gamma Repeaters (SGRs), Anomalous X-Ray pulsars (AXPs), and X-ray
dim isolated neutron stars (XDINs), and that SGRs, AXPs, and XDINs may
be linked ancestrally \cite{niebergal10:a}.

The conversion of a neutron star to a hypothetical quark star could
lead to quark novae. Such events could explain gamma ray
bursts \cite{staff08:a}, the production of heavy elements such as
platinum through r-process nucleosynthesis \cite{jaikumar07:a}, and
double-humped super-luminous supernovae \cite{ouyed13:a}.

\section{Neutron stars}

\subsection{Non-spherical neutron stars}

\begin{figure}[b]
  \hfill
  \begin{minipage}[t]{.48\textwidth}
    \begin{center}
     \includegraphics[scale=0.5]{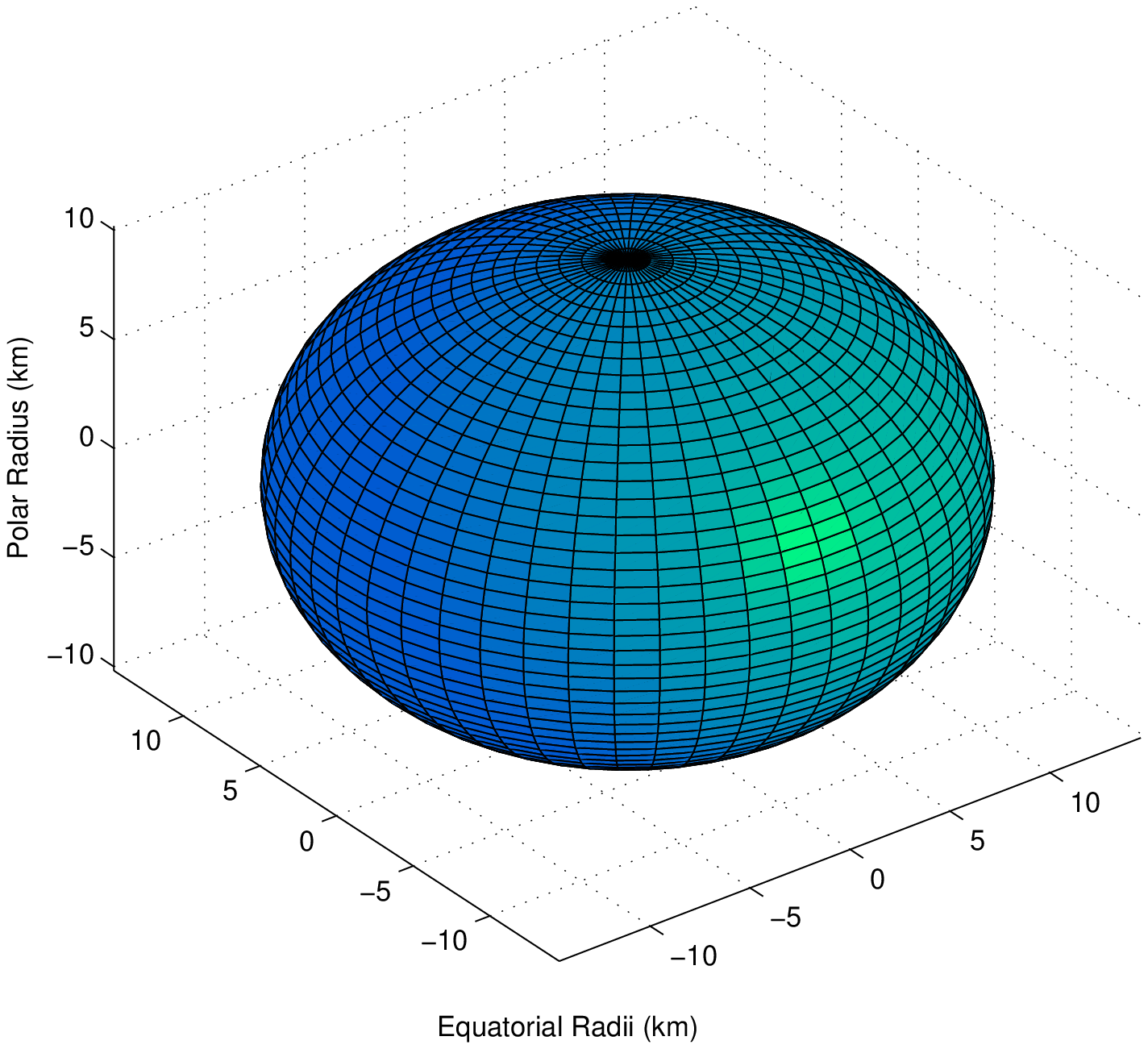}
      \caption{Oblate spheroid ($\gamma=0.70$)}
      \label{fig:g070}
    \end{center}
  \end{minipage}
  \hfill
  \begin{minipage}[t]{.48\textwidth}
    \begin{center}
     \includegraphics[scale=0.5]{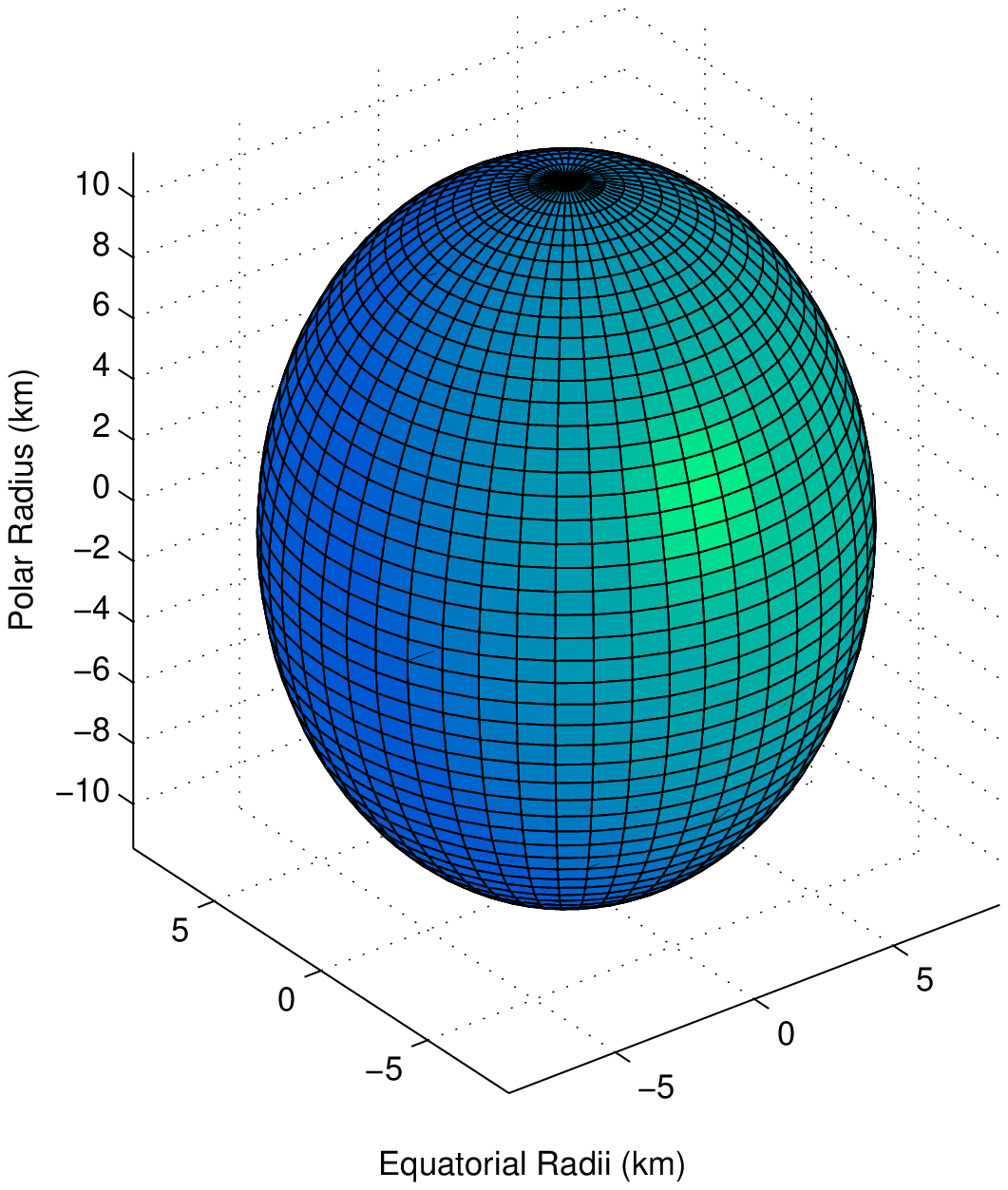}
      \caption{Prolate spheroid ($\gamma=1.30$)}
      \label{fig:g130}
    \end{center}
  \end{minipage}
  \hfill
\end{figure}
Usually, the structure of neutron stars is modeled with the assumption
that they are perfect spheres. However, due to to very high magnetic
fields, certain classes of neutron stars, such as magnetars and
neutron stars containing cores of color superconducting quark
matter \cite{ferrer10:a}, may be deformed. The stellar structure
equation of such stars is given by \cite{zubairi14:a}
\begin{eqnarray}
 \frac{dP}{dr}=-(\epsilon+P) \left(\frac{1}{2}r+4{\pi}r^{3}P
 -\frac{1}{2}r\left(1-\frac{2m}{r}\right)^{\gamma}\right)
 \left(r^{2}\left(1-\frac{2m}{r}\right)^{\gamma}\right)^{-1}~ .
\label{eq:gtov}
\end{eqnarray}
Here, $\epsilon$ is the energy-density, $P$ is the pressure, $m$ is
the gravitational mass, and $\gamma$ is a deformation constant. The
star's total mass is given by $M= \gamma m$.  Oblate (prolate) neutron
stars are obtained for $\gamma<1$ ($\gamma>1)$, as shown in
Figs.\ \ref{fig:g070} and \ref{fig:g130}, respectively.  For
$\gamma=1$, equation (\ref{eq:gtov}) reduces to the well known
Tolman-Oppenheimer-Volkoff equation, which describes the properties of
spherically symmetric neutron (compact)
stars \cite{glen97:book,frido99:book,weber05:a}.

\begin{figure}[t]
\centering
\includegraphics[width=0.5 \textwidth]{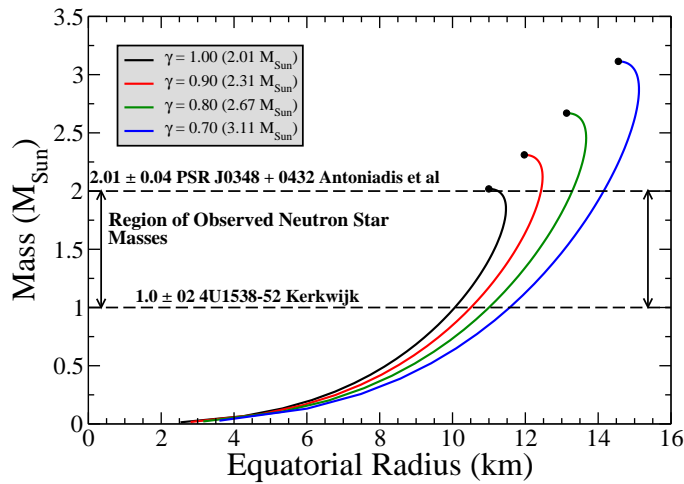}
\caption{(Color online) Mass-radius relationships of sequences of
  oblate strange quark stars. The mass is increasing with increasing
  oblateness (decreasing values of $\gamma$).  The range of observed
  neutron star masses is indicated.}
\label{fig:mr.oblate}
\end{figure}

The mass-radius relationship of oblate compact stars, computed from
Eq.\ (\ref{eq:gtov}) for different $\gamma$ values, are shown in
Fig.\ \ref{fig:mr.oblate}.

For simplicity, the equation of state of a relativistic gas of
deconfined quarks described by the MIT bag model, $P= (\epsilon -
4B)/3$, has been used here \cite{zubairi14:a}. The value of the bag
constant is $B=57~{\rm MeV}/{\rm fm}^3$ ($B^{1/4}=145$~MeV), which
makes the quark gas absolutely stable with respect to nuclear matter.
The results shown in Figs.\ \ref{fig:g070} to \ref{fig:mr.oblate}
therefore are for strange quark stars. One sees that already
relatively small deviations from spherical symmetry increase the
masses of strange stars substantially.

\subsection{Rotation-driven compositional changes}

The change in central density of a neutron star whose frequency varies
from zero to the mass shedding (Kepler) frequency can be as large as
50 to 60\% \cite{weber05:a}. This suggests that changes in the rotation
rate of a neutron star may drive phase transitions and/or lead to
significant compositional changes in the star's
core \cite{weber05:a,weber07:a,negreiros13:PLB}. As a case in point,
for some rotating neutron stars the mass and initial rotational
frequency may be just such that the central density rises from below
to above the critical density for dissolution of baryons into their
quark constituents. This may be accompanied by a sudden shrinkage of
the neutron star, effecting the star's moment of inertia and, thus,
its spin-down behavior.  As shown in Ref.\ \cite{glen97:a}, the
spin-down of such a neutron star may be stopped or even reversed for
tens of thousands to hundreds of thousands of
years \cite{weber05:a,glen97:a}.  The observation of an isolated
neutron star which is spinning-up, rather than down, could thus hint
at the existence of quark matter in its core.

\subsection{Quark deconfinement in high-mass neutron stars}

Quark deconfinement in high-mass neutron stars has very recently been
studied using extensions of the local and non-local 3-flavor
Nambu-Jona Lasinio (NJL) model supplemented with repulsive vector
interactions among the quarks (see
Refs.\ \cite{orsaria13:a,orsaria14:a} and references therein). The
phase transition from confined hadronic matter to deconfined quark
matter has been constructed via the Gibbs
\begin{figure}[htb]
\centering
\includegraphics[width=0.7 \textwidth]{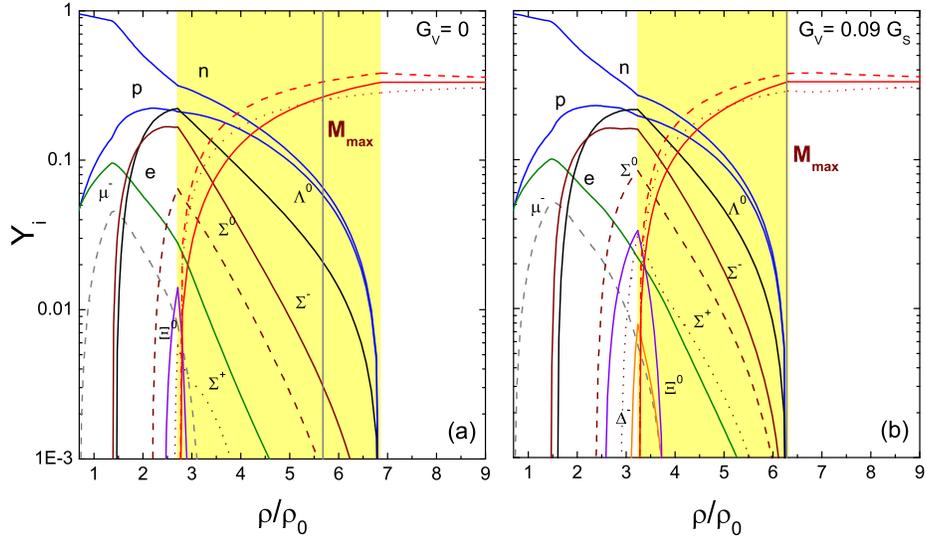}
\caption{(Color online) Particle population of neutron star matter
  computed for the non-local SU(3) NJL model. The yellow areas
  highlight the mixed phase. The solid vertical lines indicate the
  central densities of the associated maximum-mass NSs. The hadronic
  model parametrization is NL3 and the vector repulsions are (a)
  $G_V/G_S =0$ and (b) $G_V/G_S =0.09$.}
\label{pb}
\end{figure}
condition, which imposes global rather than local electric charge
neutrality and baryon number conservation. Depending on the strength
of the quark vector repulsion, it was found that an extended mixed
phase of confined hadronic matter and deconfined quarks can exist in
neutron stars as massive as $2.1$ to $2.4\, M_\odot$. A phase of pure
quark matter inside such high-mass neutron stars, while not excluded,
is only obtained for certain parametrizations of the underlying
lagrangians.  The radii of all these stars are between 12 and 13~km,
as expected for neutron stars of that mass
 \cite{Steiner10:a,Steiner:2012xt,Lattimer:2013hma}.

In Fig.\ \ref{pb} we show the relative particle fractions $Y_i$
$(\equiv \rho_i/\rho)$ of neutron star matter as a function of baryon
number density for the non-local NJL model and using NL3
parametrization of Ref.\ \cite{Lalazissis} for the hadronic model.
It can be seen that by increasing the strength of the vector
interaction, negatively charged particles like $\mu^{-}$'s and
$\Delta^{-}$'s take on the role of electrons, whose primary duty is to
make the stellar matter electrically neutral. Because of the early
onset of the $\Delta$ population
\begin{figure}[htb]
\centering
\includegraphics[width=0.6 \textwidth]{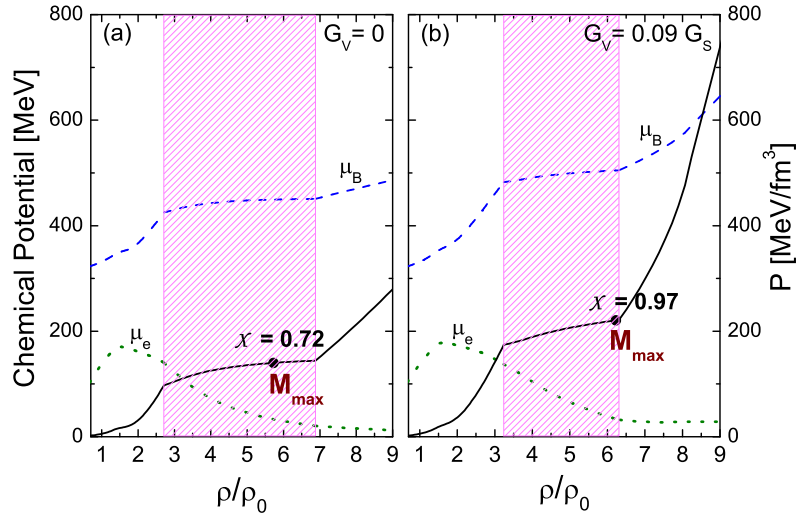}
\caption{(Color online) Pressure $P$ (solid lines), baryon chemical
  potential $\mu_B = \mu_n/3$ (dashed lines), and electron chemical
  potential $\mu_e$ (dotted lines) as a function of baryon number
  density (in units of $\rho_0 = 0.16 \, {\rm fm}^{-3}$).  In all the
  cases a non local SU(3) NJL and non linear Walecka (with
  parametrization NL3) models are considered for quark matter and
  hadronic phases, respectively. The hatched areas denote the mixed
  phase regions where confined hadronic matter and deconfined quark
  matter coexist. Panel (a) is computed for zero vector repulsion. The
  impact of finite values of the vector repulsion ($G_V=0.09 G_S$) on
  the data is shown in panel (b).}
\label{press}
\end{figure}
in this model, there is less need for electrons so that their number
density in the mixed phase is reduced compared to the outcome of
standard mean-field/bag model calculations.  The corresponding
deleptonization densities, i.e.\  the densities beyond which leptons
are no longer present in quark-hybrid matter, depend on the ratio
$G_V/G_S$ ($\rho^{G_V=0}=4.55~\rho_0$ and
$\rho^{G_V=0.09G_S}=5.33~\rho_0$) \cite{orsaria14:a}.

Since we model the quark-hadron phase transition in three-space,
accounting for the fact that the electric and baryonic charge are
conserved for neutron star matter, the pressure varies monotonically
with the proportion of the phases in equilibrium, as shown in
Fig.\ \ref{press}. The hatched areas shown in this figure denote the
mixed phase regions where confined hadronic matter and deconfined
quark matter coexist. The quark matter contents of the maximum-mass
neutron stars computed for these equations of state are also indicated
($\chi$ values).  Pure quark matter would exist for
$\chi=1$.  Our calculations show that, in the non-local case, the
inclusion of the quark vector coupling contribution shifts the onset
of the phase transition to higher densities and narrows the width of
the mixed quark-hadron phase, when compared to the case $G_V = 0$. To
the contrary, when the quark matter phase is represented by the local
NJL model, the width of the mixed phase tends to be broader for finite
$G_V/G_S$ values \cite{orsaria14:a}.

To account for the uncertainty in the theoretical predictions of the
ratio $G_V/G_S$, we treat the vector coupling constant as a free parameter. We
observed that the non local NJL model is more sensitive to the increase of
$G_V/G_S$ than the local model. For $G_V/G_S\,>\,0.09$ we have a shift
of the onset of the quark-hadron phase transition to higher and higher
densities, preventing quark deconfinement in the cores of neutron
stars.  However, we can reach values of $G_V/G_S$ up to $0.3$ for the
local NJL case.

\subsection{``Backbending''--a possible signal of quark deconfinement}

Whether or not quark deconfinement exists in static (non-rotating)
neutron stars makes only very little difference to their properties,
such as the range of possible masses and radii, which renders the
detection of quark matter in such objects extremely complicated. This
may be strikingly different for rotating neutron stars which develop
quark matter cores in the course of spin-down.  The reason being that
such stars become more and more compressed as they spin down from high
to low frequencies.  For some rotating neutron stars the mass and
initial rotational frequency may be just such that the central density
rises from below to above the critical density for the dissolution of
baryons into their quark constituents. This could affect the star's
moment of inertia dramatically \cite{weber05:a,glen97:a}. Depending on
the rate at which quark matter is produced, the moment of inertia can
decrease very anomalously, and could even introduce an era of stellar
spin-up (so-called ``backbending'') lasting for $\sim 10^8$
years \cite{glen97:a}. Since the dipole age of millisecond pulsars is
about $10^9$~years, one may estimate that roughly about 10\% of the
solitary millisecond pulsars presently known could be in the quark
transition epoch and thus could be signaling the ongoing process of
quark deconfinement.  Changes in the moment of inertia reflect
themselves in the braking index, $n$, of a rotating neutron star, as
can be seen from \cite{weber05:a,glen97:a,spyrou02:a}
\begin{equation}
  n(\Omega) \equiv \frac{\Omega\, \ddot{\Omega} }{\dot{\Omega}^2} = 3
  - \frac{ I + 3 \, I' \, \Omega + I'' \, \Omega^2 } {I + I' \,
  \Omega} \rightarrow 3 - \frac{ 3 \, I' \, \Omega + I'' \, \Omega^2
  } {2\, I + I' \, \Omega}
\label{eq:index}
\end{equation}
where dots and primes denote derivatives with respect to time and
$\Omega$, respectively.  The last relation in Eq.\ (\ref{eq:index})
constitutes the non-relativistic limit of the braking index. It is
obvious that these expressions reduce to the canonical limit, $n=3$,
if the moment of inertia is completely independent of
frequency. Evidently, this is not the case for rapidly rotating
neutron stars, and it fails for stars that experience pronounced
internal changes (as possibly driven by phase transitions) which alter
the moment of inertia significantly. In Ref.\ \cite{glen95:a} it
was shown that the changes in the moment of inertia caused by the
gradual transformation of hadronic matter into quark matter may lead
to $n(\Omega) \rightarrow \pm \infty$ at the transition frequency
where pure quark matter is produced.  Such dramatic anomalies in
$n(\Omega)$ are not known for conventional neutron stars (see,
however, Ref.\ \cite{zdunik05:a}), because their moments of inertia
appear to vary smoothly with $\Omega$ \cite{frido99:book}. The future
astrophysical observation of a strong anomaly in the braking behavior
of a pulsar may thus indicate that quark deconfinement is occurring at
the pulsar's center.

\subsection{Quark-hadron Coulomb lattices in the cores of neutron stars}

\begin{figure}
\centering
  \includegraphics[scale=0.32]{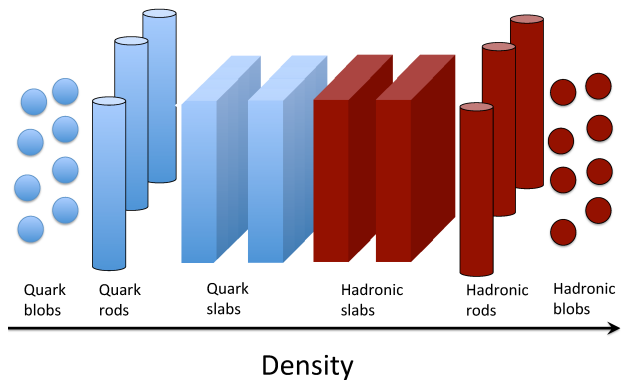}
  \caption{Schematic illustration of possible geometrical structures
    in the quark-hadron mixed phase of neutron stars. The structures
    may form because of the competition between the Coulomb and the
    surface energies associated with the positively charged regions of
    nuclear matter and negatively charged regions of quark matter.}
\label{fig:qh_structures}
\end{figure}

Because of the competition between the Coulomb and the surface
energies associated with the positively charged regions of nuclear
matter and negatively charged regions of quark matter, the mixed phase
may develop geometrical structures (e.g., blobs, rods, slabs, as
schematically illustrated in
Fig.\ \ref{fig:qh_structures} \cite{na12:a}), similarly to what is
expected of the sub-nuclear liquid-gas phase
transition \cite{ravenhall83:a,ravenhall83:b,williams85:a,glendenning92:a}.
The consequences of such a Coulomb lattice for the thermal and
transport properties of neutron stars have been studied in
Ref.\ \cite{na12:a}.  It was found that at low
temperatures of $T\lesssim 10^8$~K the neutrino emissivity from
electron-blob Bremsstrahlung scattering is at least as important as
the total contribution from all other Bremsstrahlung processes (such
as nucleon-nucleon and quark-quark Bremsstrahlung) and modified
nucleon and quark Urca processes.  It is also worth noting that the
scattering of degenerate electrons off rare phase blobs in the mixed
phase region lowers the thermal conductivity by several orders of
magnitude compared to a quark-hadron phase without geometric patterns.
This may lead to significant changes in the thermal evolution of the
neutron stars containing solid quark-hadron cores, which has not yet
been studied.

\subsection{Pycnonuclear reaction rates}

The presence of strange quark nuggets in the crustal matter of neutron
stars could be a consequence of Witten's strange quark matter
hypothesis \cite{witten84:a}. The impact of such nuggets on the
pycnonuclear reaction rates among heavy atomic nuclei has been studied
in Ref.\ \cite{golf09:a}.  Particular emphasis was put on the
consequences of color superconductivity on the reaction rates.
Depending on whether or not quark nuggets are in a color
superconducting state, their electric charge distributions differ
drastically, which was found to have dramatic consequences for the
pycnonuclear reaction rates in the crusts of neutron stars. Future
nuclear fusion network calculations may thus have the potential to
shed light on the existence of strange quark matter nuggets and on
whether they are in a color superconducting state, as suggested by
QCD.

\subsection{Rotational instabilities}

The r-mode instability of a rotating neutron star dissipates the
star's rotational energy by coupling the angular momentum of the star
to gravitational
waves \cite{andersson98:a,lindblom98:a,friedman98:a}. This instability
can be active in a newly formed isolated neutron star as well as in
old neutron stars being spun up by accretion of matter from binary
stars. If the interior contains quark matter, the r-mode instability
and the gravitational wave signal may carry information about quark
matter \cite{jaikumar08:a,rupak1013:a,rupak1013:b,sad08:a,mannarelli07:a}.

\section*{Acknowledgments}

F.\ W.\ is supported by the National Science Foundation (USA) under
Grants PHY-0854699 and PHY-1411708.  M.\ O.\ and G.\ C.\ acknowledge CONICET and
SeCyT-UNLP (Argentina) for financial support. M.\ O.\ and G.\ C.\ are
thankful for hospitality extended to them during their visits at the
SDSU, supported by a NSF-CONICET International Cooperation Project.

\end{document}